\documentclass[times, 10pt,onecolumn]{article}
\usepackage{amssymb}
\usepackage{amsmath}
\usepackage{latex8}
\usepackage{times}

\input{epsf}        
\newcommand{\cepsffig}[1]{\begin{center}{\mbox{\epsffile{#1}}}\end{center}}

\begin{document}

\title{Mining Cellular Automata DataBases throug PCA Models}
\author{Gilson A. Giraldi \\
LNCC--National Laboratory of Scientific Computation \\
Av. Getulio Vargas, 333, 25651-070 Rio de Janeiro, \\
RJ, Brazil\\
gilson@lncc.br \\
\\
Antonio A.F. Oliveira,\quad Leonardo Carvalho \\
LCG--Computer Graphics Laboratory, UFRJ-COPPE \\
Mail Box 68511, 21945-970 Rio de Janeiro, RJ, Brazil \\
\{oliveira,leonardo\}@lcg.ufrj.br}
\maketitle

\begin{abstract}
Cellular Automata are discrete dynamical systems that evolve following
simple and local rules. Despite of its local simplicity, knowledge discovery
in CA is a NP problem. This is the main motivation for using data mining
techniques for CA study. The Principal Component Analysis (PCA) is a useful
tool for data mining because it provides a compact and optimal description
of data sets. Such feature have been explored to compute the best subspace
which maximizes the projection of the I/O patterns of CA onto the principal
axis. The stability of the principal components against the input patterns
is the main result of this approach. In this paper we perform such analysis
but in the presence of noise which randomly reverses the CA output values
with probability $p$. As expected, the number of principal components
increases when the pattern size is increased. However, it seems to remain
stable when the pattern size is unchanged but the noise intensity gets
larger. We describe our experiments and point out further works using KL
transform theory and parameter sensitivity analysis.
\end{abstract}

\thispagestyle{empty}

\Section{Introduction}

Data Mining is a shorter term that refers to extracting or \textit{mining}
knowledge (\textit{interesting patterns}) from large amounts of data \cite%
{han-Kamber}. The techniques in this field varies from computational methods
(Decision Trees, Data Structures, etc.) statistical and regression
(Correlation, Linear and Logistic Regression, Cluster Analysis,etc.), neural
networks, and dimension reduction (PCA and Singular Value Decomposition ) 
\cite{han-Kamber}.

When applied for scientific data sets, such patterns lead to conjectures
about system behavior and properties which must be analyzed through a
theoretical framework in order to confirm its truth. That is the philosophy
of this work. Following \cite{deniau94pca}, we apply PCA methods for
Cellular Automata Analysis.

Cellular automata are discrete dynamical systems \cite{DGT83,demongeot85}
originally proposed by Von Neumann \cite{Neumann1966} (see also \cite{349202}
for a brief story). They consist of a lattice of discrete identical sites,
each site taking on a finite set of values \cite{compref347,513738}. The
values of the sites evolve in discrete time steps according to simple rules
that update the value of each site in terms of the values of neighboring
sites \cite{513738,wolfram:94}. Cellular Automata is a rich field of
investigation that includes computational aspects like Universality,
languages/grammars and state transition diagrams \cite{wolfram84b,compref347}%
, statistical mechanics and probability (self-organization, Markov theory,
fractals, etc.) \cite%
{wolfram:94,gutowitz89c,schulman78,wolfram83,gutowitz90b}, algebraic methods
(matrix algebra, polynomials over Finite Fields, etc.) \cite%
{martin84,wolfram:94,pedersen92,sieburg91,146255,626727,Chaudhuri1997} among
others \cite{513738,boccara91}. They have been applied for pattern
classification and recognition \cite{morales:98,gutowitz89b,gutowitz89c},
pattern generation \cite{bays87b,boccara91,626727}, hardware architectures
for massively parallel computation \cite{cam8,gutowitz93b}, models for
biological process \cite{bard:93,ermentrout93,hartman00} and physical
systems simulation \cite%
{fang:02b,Kadanoff:89,Pires:89,wolfram:86,Succi2002,BC-livre,chopard:02c}.

Despite of its local simplicity, knowledge discovery in CA is a NP problem 
\cite{wolfram:94}. This is the main motivation for using data mining
techniques for CA study. For instance, in \cite{deniau94pca} a set of binary
one-dimensional cellular automata is considered. Each such CA is feed with a
set of input patterns and the obtained output data base is analyzed through
Principal Component Analysis.

In this paper we follow such viewpoint but in the presence of noise which
randomly reverses the CA output values with probability $p$. As expected,
the number of principal components increases when the pattern size is
increased. However, it seems to remain stable when the pattern size is
unchanged but the noise intensity gets larger. We describe our experiments
and point out further works using KL transform theory and parameter
sensitivity analysis.

This paper is organized as follows. The next section presents the basic
concepts of CAs and how computational intractable problems arise in this
area. Next, in section \ref{PCA}, we discuss the Principal Component
Analysis. The application of PCA for cellular automata analysis is reviewed
in section \ref{PCA-Cell}. Finally, we present our results and final
comments on sections \ref{PCA-ProbCell} and \ref{Final}, respectively.

\Section{Cellular Automata \label{CA}}

A cellular automaton (CA) is a quadruplet $A=(L;S;N;f)$ where $L$ is a set
of indices or sites, $S$ is the finite set of site values or states, $%
N:L\rightarrow L^{k}$ is a one-to-many mapping defining the neighborhood of
every site $i$ as a collection of $k$ sites, and $f:S^{k}\rightarrow S$ is
the evolution function of $A$ \cite{wolfram:94,Adami1998}. The neighborhood
of site $i$ is defined as the set $N(i)=\{j;|j-i|\leq [(k-1)/2]\}$ ($[x]$
stands for the integer part of $x$); one must notice that a given site may
or not be included in its own neighborhood. Since the set of states is
finite, $\{f_{j}\}$ will denote the set of possible rules of the CA taken
among the $p=(\#S)^{(\#S)^{k}}$ rules.

\begin{figure}[tbph]
\epsfxsize=6.0cm
\par
\begin{center}
{\mbox{\epsffile{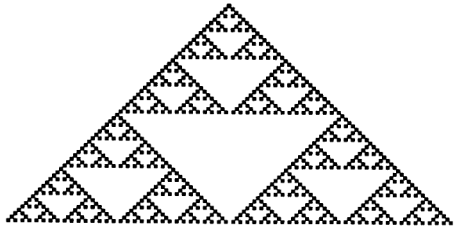}}}
\end{center}
\caption{Evolution of CA given by rule number $90$. In this case, the
initial configuratio is a finite one-dimensional lattice which has only one
site with the value $1$ (pictured in black). 
Source: www.stephenwolfram.com/publications/articles/ca/}
\label{rule-90}
\end{figure}

For a one-dimensional cellular automaton the lattice $L$ is an array of
sites, and the transition rule $f$ updates a site value according to the
values of a neighborhood of $k=2r+1$ sites around it, that means:

\begin{equation}
f:S^{2r+1}\rightarrow S,  \label{ca001}
\end{equation}

\begin{eqnarray}
a_{i}^{t+1} &=&f\left(
a_{i-r}^{t},...,a_{i-1}^{t},a_{i}^{t},a_{i+1}^{t},...,a_{i+r}^{t}\right) ,
\label{ca01} \\
a_{j}^{t} &\in &S,\quad j=i-r,...,i+r.
\end{eqnarray}
where $t$ means the evolution time, also taking discrete values, and $%
a_{i}^{t}$ means the value of the site $i$ at time $t$ \cite%
{wolfram:94,513738} (see also \cite{WolframSite} for on-line examples).
Therefore, given a configuration of site values at time $t$, it will be
updated through the application of the transition rule to generate the new
configuration at time $t+1$, and so on. In the case of $r=1$ in expression (%
\ref{ca01}) and $S=\left\{ 0,1\right\} $ we have a special class of cellular
automata which was extensively studied in the CA literature \cite%
{349202,626727,Chaudhuri1997,wolfram:94}. Figure \ref{rule-90} shows the
very known example of such a CA. The rule in this case is:

\begin{equation}
a_{i}^{t+1}=\left( a_{i-1}^{t}+a_{i+1}^{t}\right) mod2,  \label{ca02}
\end{equation}
that means, the remainder of the division by two. The 
figure pictures the
evolution of an initial configuration in which there is only one site with
the value $1$.

Once $r=1$ in expression (\ref{ca01}), it is easy to check that this rule is
defined by the function:

\begin{equation}
\begin{array}{lll}
1 & 1 & 1 \\ 
& 0 & 
\end{array}%
\quad 
\begin{array}{lll}
1 & 1 & 0 \\ 
& 1 & 
\end{array}%
\quad 
\begin{array}{lll}
1 & 0 & 1 \\ 
& 0 & 
\end{array}%
\quad 
\begin{array}{lll}
1 & 0 & 0 \\ 
& 1 & 
\end{array}%
\begin{array}{lll}
0 & 1 & 1 \\ 
& 1 & 
\end{array}%
\quad 
\begin{array}{lll}
0 & 1 & 0 \\ 
& 0 & 
\end{array}%
\quad 
\begin{array}{lll}
0 & 0 & 1 \\ 
& 1 & 
\end{array}%
\quad 
\begin{array}{lll}
0 & 0 & 0 \\ 
& 0 & 
\end{array}
\label{regra-90}
\end{equation}

\begin{equation}
0\ast 2^{7}+1\ast 2^{6}+0\ast 2^{7}+1\ast 2^{4}+1\ast 2^{3}+0\ast
2^{2}+1\ast 2^{1}+0\ast 2^{0}=90  \label{regra-90-idex}
\end{equation}

\begin{figure}[htbp]
\epsfxsize=12.0cm
\par
\begin{center}
\begin{minipage}[b]{6.0cm}
    \begin{center}
      \epsfxsize=6.0cm
      \cepsffig{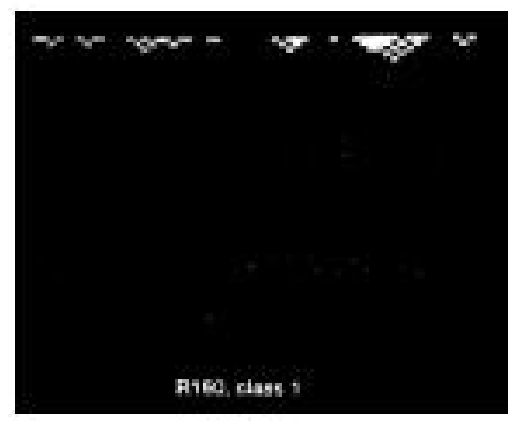}
    (a)
    \end{center}
  \end{minipage}
\begin{minipage}[b]{6.0cm}
    \begin{center}
      \epsfxsize=6.0cm
      \cepsffig{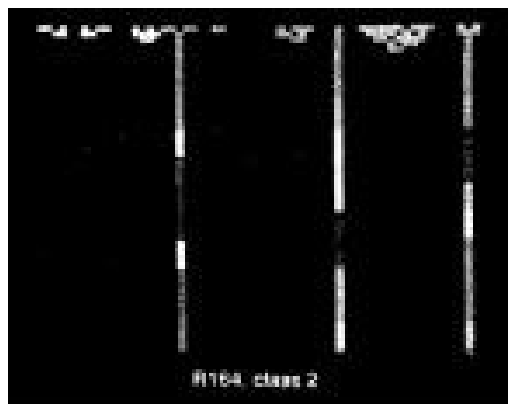}
    (b)
    \end{center}
  \end{minipage}
\par
\begin{minipage}[b]{6.0cm}
    \begin{center}
      \epsfxsize=6.0cm
      \cepsffig{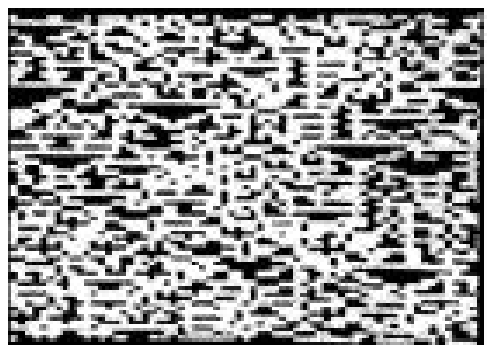}
    (c)
    \end{center}
  \end{minipage}
\begin{minipage}[b]{6.0cm}
    \begin{center}
      \epsfxsize=6.0cm
      \cepsffig{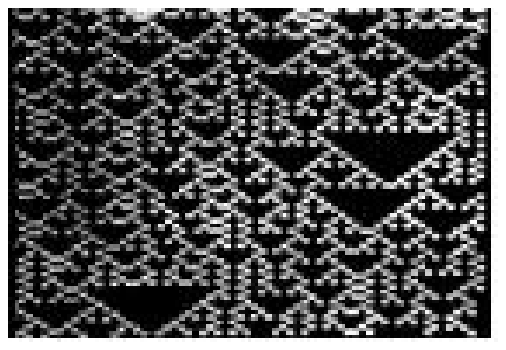}
    (d)
    \end{center}
  \end{minipage}
\end{center}
\caption{Some examples of Wolfram's classification for one-dimensional $r=1$
CAs. Source: www.stephenwolfram.com/publications/articles/ca/}
\label{classes}
\end{figure}

By observing this example, we see that there are $2^{8}=256$ such rules and
for each one it can be assigned a \textit{rule number} following the
indexation illustrated on expression \ref{regra-90-idex}. In \cite{wolfram84b}%
, Wolfram proposes four basic classes of behavior for these rules (see also 
\cite{Adami1998}):

Class 1: Evolution leads to homogeneous state in which all the sites have
the same value (Figure \ref{classes}.a);

Class 2: Evolution leads to a set of stable and periodic structures that are
separated and simple (Figure \ref{classes}.b);

Class 3: Evolution leads to a chaotic pattern (Figure \ref{classes}.c);

Class 3: Evolution leads to complex structures (Figure \ref{classes}.d).

Other classifications based on Markovian processes and group properties can
be also found in the literature \cite{gutowitz90b,146255}.

Despite of its local simplicity, knowledge discovery in CA is a NP problem.
In fact, let us take a one-dimensional CA with a finite lattice $L$ of size $%
d$. One may consider the question of whether a particular sequence of $d$
site values can occur after $T$ time steps in the evolution of the cellular
automaton, starting from any initial state. Then, one may ask whether there
exists any algorithm that can determine the answer in a time given by some
polynomial in $d$ and $T$. The question can certainly be answered by testing
all sequences of possible initial site values, that is $(\#S)^{d}$. But this
procedure requires a time that grows exponentially with $d$.

Nevertheless, if an initial sequence could be guessed, then it could be
tested in a time polynomial in $d$ and $T$. As a consequence, the problem is
in the class NP which motivates the application of data mining techniques
for knowledge discovery in CA. The next sections review PCA basic theory and
its application for the analysis of the (traditional) set of rules composed
by $256$ $1D$ cellular automata obtained when $r=1$, $S=\left\{ 0,1\right\} $%
.

\Section{Principal Component Analysis \label{PCA}}

\begin{figure}[htbp]
\epsfxsize=10.0cm
\par
\begin{center}
\begin{minipage}[b]{5.0cm} 
    \begin{center} 
      \epsfxsize=5.0cm 
      \cepsffig{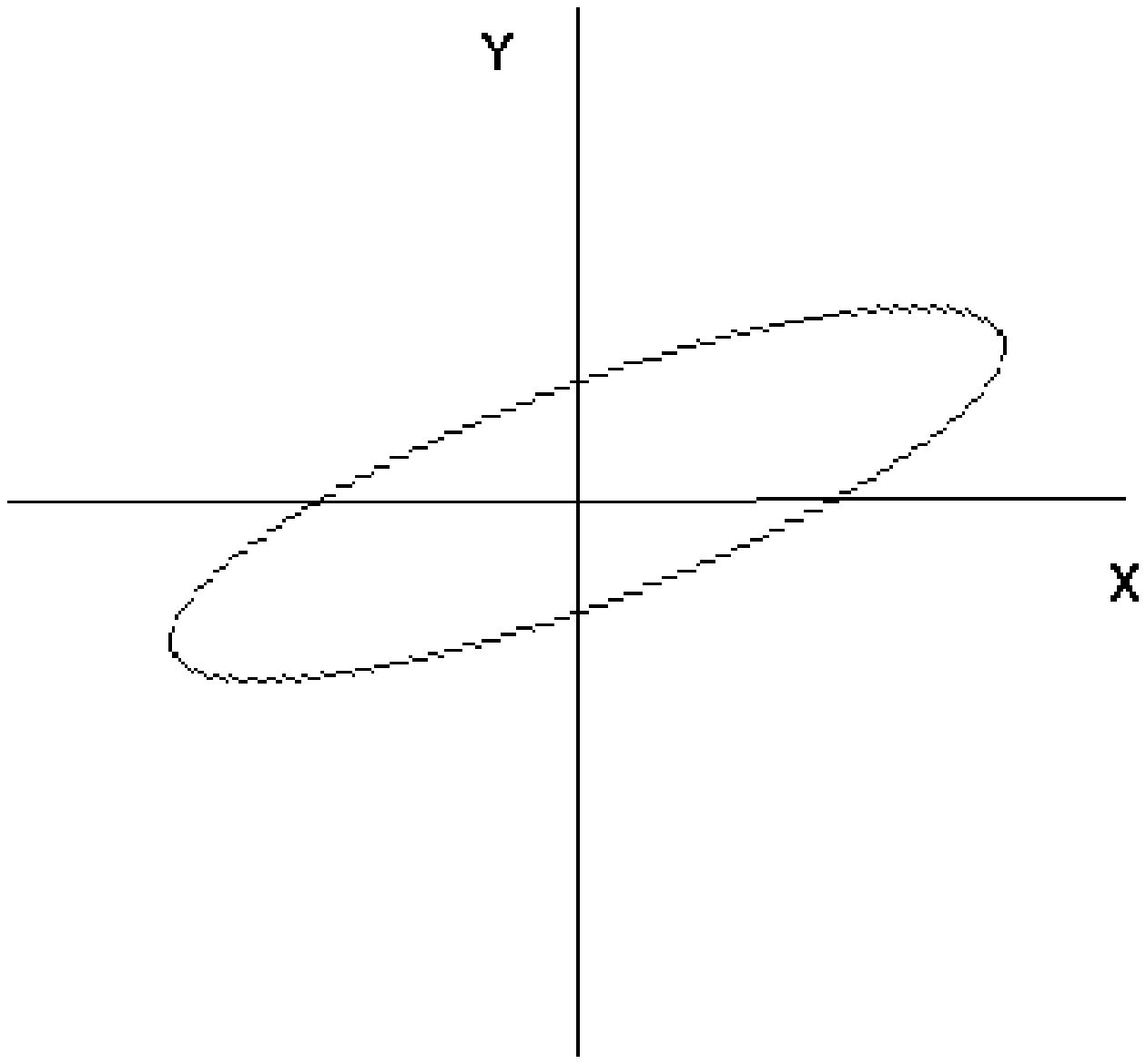} 
    (a) 
    \end{center} 
  \end{minipage} 
\begin{minipage}[b]{5.0cm} 
    \begin{center} 
      \epsfxsize=5.0cm 
      \cepsffig{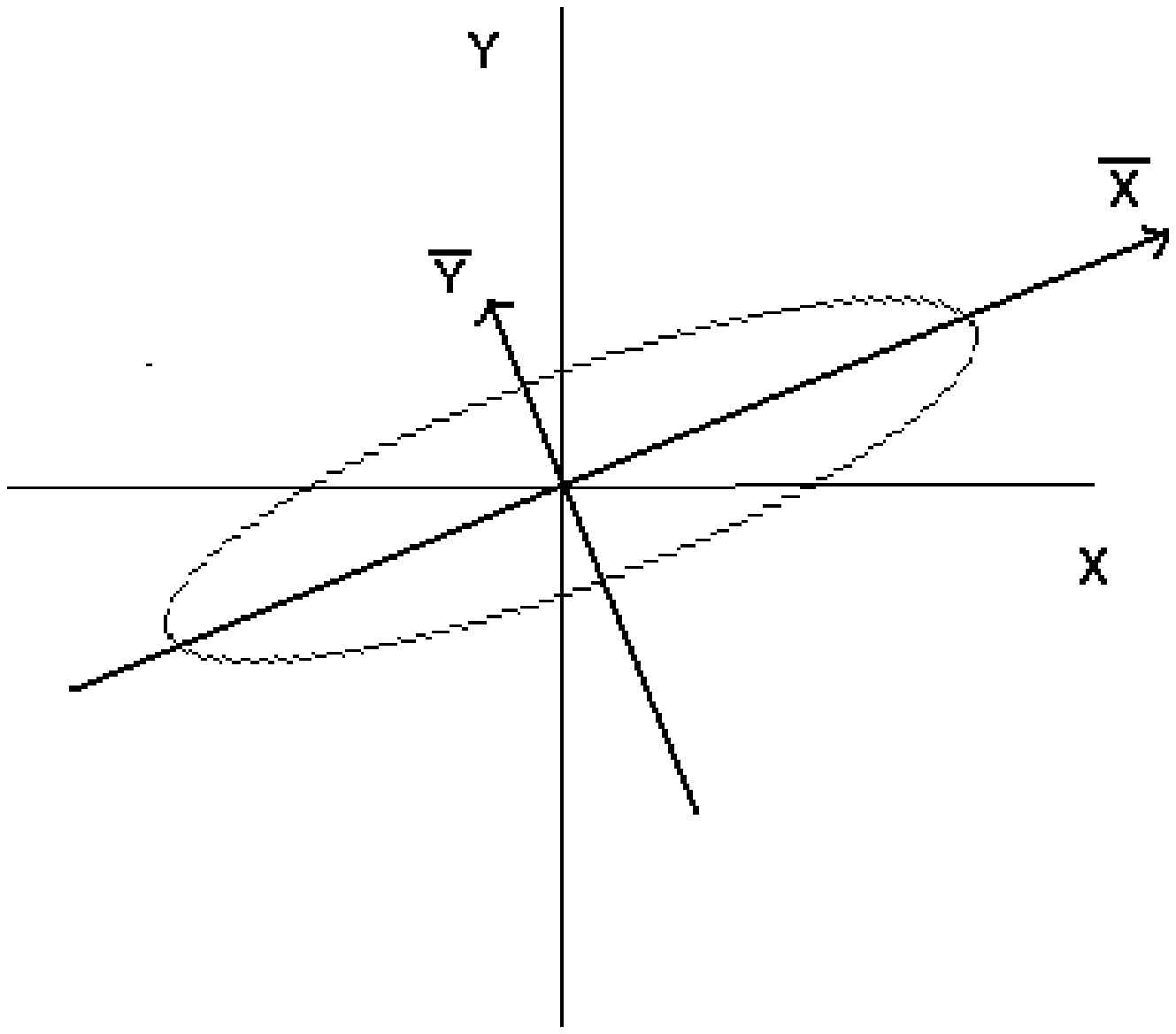} 
    (b) 
    \end{center} 
  \end{minipage}
\end{center}
\caption{(a)Original dataset. (b) Extraction of the principal component.}
\label{data}
\end{figure}

Principal Component Analysis (\textbf{PCA}), also called Karhunen-Loeve, or
KL method, can be seen as a method for data compression or dimensionality
reduction \cite{Algazi1969} (see \cite{Jain89}, section $5.11$ also). Thus,
let us suppose that the data to be compressed consist of $N$ tuples or data
vectors, from a n-dimensional space. Then, PCA
searches for $k$ n-dimensional orthonormal vectors that can best be used to
represent the data, where $k\leq n$. Figure \ref{data}.a,b picture this idea
using a bidimensional representation. If we suppose the the data points are distributed
over the elipse, it follows that the coordinate system ($\left( \overline{X},\overline{Y}\right) $, 
shown in Figure \ref{data}.b
is more suitable for representing the data set in a sense that will be 
formally described next.

Thus, let $S=\left\{ \mathbf{u}_{1},\mathbf{u}_{2},...,\mathbf{u}%
_{N}\right\} $ the data set represented on Figure \ref{data}. By now, let us
suppose that the centroid of the data set is the center of the coordinate
system, that means:

\begin{equation}
C_{M}=\frac{1}{N}\sum_{i=1}^{N}\mathbf{u}_{i}=\mathbf{0}.  \label{centroid00}
\end{equation}

To address the issue of compression, we need a vector basis that satisfies a
proper optimization criterium (rotated axes in Figure \ref{data}.b).
Following \cite{Jain89}, consider the operations in Figure \ref{KL}. The
vector $\mathbf{u}_{j}$ is first transformed to a vector $\mathbf{v}_{j}$ by
the matrix (transformation) $A$. Thus, we truncate $\mathbf{v}_{j}$ by
choosing the first $m$ elements of $\mathbf{v}_{j}$. The obtained vector $%
\mathbf{w}_{j}$ is just the transformation of $\mathbf{v}_{j}$ by $I_{m}$,
that is a matrix with 1s along the first $m$ diagonal elements and zeros
elsewhere. Finally, $\mathbf{w}_{j}$ is transformed to $\mathbf{z}_{j}$ by
the matrix $B$. Let the square error defined as follows:

\begin{equation}
J_{m}=\frac{1}{N}\sum_{j=0}^{N}\left\| \mathbf{u}_{j}-\mathbf{z}_{j}\right\|
^{2}=\frac{1}{n}Tr\left[ \sum_{j=0}^{N}\left( \mathbf{u}_{j}-\mathbf{z}%
_{j}\right) \left( \mathbf{u}_{j}-\mathbf{z}_{j}\right) ^{*T}\right] ,
\label{ca04}
\end{equation}
where $Tr$ means the trace of the matrix between the square brackets and the
notation ($*T $) means the transpose of the complex conjugate of a matrix.
Following Figure \ref{KL}, we observe that $\mathbf{z}_{j}=BI_{m}A\mathbf{u}%
_{j}$. Thus we can rewrite (\ref{ca04}) as:

\begin{figure}[tbph]
\epsfxsize=10.0cm
\par
\begin{center}
{\mbox{\epsffile{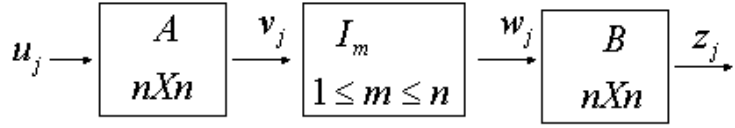}}}
\end{center}
\caption{KL Transform formulation. Reprinted from \cite{Jain89}.}
\label{KL}
\end{figure}

\begin{equation}
J_{m}=\frac{1}{N}Tr\left[ \sum_{i=0}^{N}\left( \mathbf{u}_{j}-BI_{m}A\mathbf{%
u}_{j}\right) \left( \mathbf{u}_{j}-BI_{m}A\mathbf{u}_{j}\right) ^{*T}\right]
,  \label{ca05}
\end{equation}
which yields:

\begin{equation}
J_{m}=\frac{1}{N}Tr\left[ \left( I-BI_{m}A\right) R\left( I-BI_{m}A\right)
^{*T}\right] ,  \label{ca06}
\end{equation}
where:

\begin{equation}
R=\sum_{i=0}^{N}\mathbf{u}_{j}\mathbf{u}_{j}^{*T}.  \label{ca07}
\end{equation}

Following the literature, we call $R$ the covariance matrix. We can now
stating the optimization problem by saying that we want to find out the
matrices $A,B$ that minimizes $J_{m}$. The next theorem gives the solution
for this problem.

\textit{Theorem 1: }The error $J_{m}$ in expression (\ref{ca06}) is minimum
when

\begin{equation}
A=\Phi ^{*T},\quad B=\Phi ,\quad AB=BA=I,
\end{equation}
where $\Phi $ is the matrix obtained by the orthonormalized eigenvectors of $%
R$ arranged according to the decreasing order of its eigenvalues.

\textit{Proof.} To minimize $J_{m}$ we first observe that $J_{m}$ must be
zero if $m=n.$ Thus, the only possibility would be

\begin{equation}
I=BA\Rightarrow A=B^{-1}.  \label{ca08}
\end{equation}

Besides, by remembering that

\begin{equation}
Tr\left( CD\right) =Tr\left( DC\right) ,  \label{prop00}
\end{equation}

we can also write:

\begin{equation}
J_{m}=\frac{1}{n}Tr\left[ \left( I-BI_{m}A\right) ^{*T}\left(
I-BI_{m}A\right) R\right] .  \label{ca09}
\end{equation}

Again, this expression must be null if $m=n$. Thus:

\begin{equation*}
J_{n}=\frac{1}{n}Tr\left[ \left( I-BA-A^{*T}B^{*T}+A^{*T}B^{*T}BA\right) R%
\right] .
\end{equation*}

This error is minimum if:

\begin{equation}
B^{*T}B=I,\quad A^{*T}A=I,  \label{ca10}
\end{equation}
that is, if $A$ and $B$ are unitary matrix. The next condition comes from
the differentiation of $J_{m}$ respect to the elements of $A$. We should set
the result to zero in order to obtain the necessary condition to minimize $%
J_{m}$. This yields:

\begin{equation}
I_{m}A^{*T}\left( I-A^{*T}I_{m}A\right) R=0,  \label{ca11}
\end{equation}
which renders:

\begin{equation}
J_{m}=\frac{1}{n}Tr\left[ \left( I-A^{*T}I_{m}A\right) R\right] .
\label{ca12}
\end{equation}

By using the property (\ref{prop00}), the last expression can be rewritten as

\begin{equation*}
J_{m}=\frac{1}{n}Tr\left[ R-I_{m}ARA^{*T}\right] .
\end{equation*}

Since $R$ is fixed, $J_{m}$ will be minimized if

\begin{equation}
\overset{\symbol{126}}{J}_{m}=Tr\left[ I_{m}ARA^{*T}\right]
=\sum_{i=0}^{m-1}a_{i}^{T}Ra_{i}^{*},  \label{ca13}
\end{equation}
is maximized where $a_{i}^{T}$ is the \textit{ith} row of $A$. Once $A$ is
unitary, we must impose the constrain:

\begin{equation}
a_{i}^{T}a_{i}^{*}=1.  \label{ca14}
\end{equation}

Thus, we shall maximize $\overset{\symbol{126}}{J}_{m}$ subjected to the
last condition. The Lagrangian has the form:

\begin{equation*}
\overset{\symbol{126}}{J}_{m}=\sum_{i=0}^{m-1}a_{i}^{T}Ra_{i}^{*}+%
\sum_{i=0}^{m-1}\lambda _{i}\left( 1-a_{i}^{T}a_{i}^{*}\right) ,
\end{equation*}
where the $\lambda _{i}$ are the Lagrangian multipliers. By differentiating
this expression respect to $a_{i}$ we get:

\begin{equation}
Ra_{i}^{*}=\lambda _{i}a_{i}^{*},  \label{ca15}
\end{equation}

Thus, $a_{i}^{*}$ are orthonormalized eigenvectors of $R$. Substituting this
result in expression (\ref{ca13}) produces:

\begin{equation}
\overset{\symbol{126}}{J}_{m}=\sum_{i=0}^{m-1}\lambda _{i},  \label{ca16}
\end{equation}
which is maximized if $\left\{ a_{i}^{*},\quad i=0,1,...,m-1\right\} $
correspond to the largest eigenvalues of $R$. ($\square $)

A straightforward variation of the above statement is obtained if we have a
random vector $\mathbf{u}$ with zero mean. In this case, the pipeline of
Figure \ref{KL} yields a random vector $\mathbf{z}$ and the square error can
be expressed as:

\begin{equation*}
J_{m}=\frac{1}{n}Tr\left[ E\left\{ \left( \mathbf{u}-BI_{m}A\mathbf{u}%
\right) \left( \mathbf{u}-BI_{m}A\mathbf{u}\right) ^{*T}\right\} \right] ,
\end{equation*}
which can be written as:

\begin{equation}
J_{m}=\frac{1}{n}Tr\left[ \left( I-BI_{m}A\right) R\left( I-BI_{m}A\right)
^{*T}\right] ,  \label{ca17}
\end{equation}
where $R=E\left( \mathbf{uu}^{*T}\right) $ is the covariance matrix.
Besides, if $C_{m}$ in expression (\ref{centroid00}) is not zero, we must
translate the coordinate system to $C_{m}$ before computing the matrix $R$ ,
that is:

\begin{equation}
\widetilde{\mathbf{u}_{j}}=\mathbf{u}_{j}\mathbf{-C}_{m}.  \label{ca18}
\end{equation}

In this case, matrix $R$ will be given by:

\begin{equation*}
R=\sum_{i=0}^{N}\widetilde{\mathbf{u}_{j}}\widetilde{\mathbf{u}_{j}}^{*T}.
\end{equation*}

Also, sometimes may be useful to consider in expression (\ref{ca04}) some
other norm, not necessarily the 2-norm. In this case, there will be a real,
symmetric and positive-defined matrix $M$, that defines the norm. Thus, the
square error $J_{m}$ will be rewritten in more general form:

\begin{equation}
J_{m}=\frac{1}{n}\sum_{j=0}^{N}\left\| \mathbf{u}_{j}-\mathbf{z}_{j}\right\|
_{M}^{2}=\frac{1}{n}\sum_{j=0}^{N}\left( \mathbf{u}_{j}-\mathbf{z}%
_{j}\right) ^{*T}M\left( \mathbf{u}_{j}-\mathbf{z}_{j}\right) .  \label{ca19}
\end{equation}

Obviously, if $M=I$ we recover expression (\ref{ca04}). The link between
this case and the above one is easily obtained by observing that there is
non-singular and real matrix $W$, such that:

\begin{equation}
W^{T}MW=I.  \label{ca20}
\end{equation}

The matrix $W$ defines the transformation:

\begin{equation}
W\widehat{\mathbf{u}_{j}}=\mathbf{u}_{j},\quad W\widehat{\mathbf{z}_{j}}=%
\mathbf{z}_{j}.  \label{ca0020}
\end{equation}

Thus, by inserting these expressions in equation (\ref{ca19}) we obtain:

\begin{equation}
J_{m}=\frac{1}{n}\sum_{j=0}^{N}\left( \widehat{\mathbf{u}_{j}}-\widehat{%
\mathbf{z}_{j}}\right) ^{*T}\left( \widehat{\mathbf{u}_{j}}-\widehat{\mathbf{%
z}_{j}}\right) .  \label{ca21}
\end{equation}

Expression (\ref{ca21}) can be written as:

\begin{equation}
J_{m}=\frac{1}{n}\sum_{j=0}^{N}\left\| \widehat{\mathbf{u}_{j}}-\widehat{%
\mathbf{z}_{j}}\right\| ^{2},  \label{ca22}
\end{equation}
now using the 2-norm, like in expression (\ref{ca04}). Therefore:

\begin{equation}
J_{m}=\frac{1}{n}Tr\left[ \sum_{j=0}^{N}\left( \widehat{\mathbf{u}_{j}}-%
\widehat{\mathbf{z}_{j}}\right) \cdot \left( \widehat{\mathbf{u}_{j}}-%
\widehat{\mathbf{z}_{j}}\right) ^{*T}\right] .  \label{ca23}
\end{equation}

Following the same development performed above, we will find that we must
solve the equation:

\begin{equation}
\widehat{R}\widehat{a_{i}^{*}}=\lambda _{i}\widehat{a_{i}^{*}},  \label{ca24}
\end{equation}
where:

\begin{equation}
\widehat{R}=\sum_{j=0}^{N}\widehat{\mathbf{u}_{j}}\widehat{\mathbf{u}_{j}}%
^{*T}.  \label{ca25}
\end{equation}

Thus, from transformations (\ref{ca0020}) it follows that:

\begin{equation}
\widehat{R}=WRW^{T}.  \label{ca26}
\end{equation}
and, therefore, we must solve the following eigenvalue/eigenvector problem:

\begin{equation}
\left( WRW^{T}\right) \widehat{a_{i}^{*}}=\lambda _{i}\widehat{a_{i}^{*}}.
\label{ca27}
\end{equation}

The eigenvectors, in the original coordinate system, are finally given by:

\begin{equation}
W\widehat{a_{i}^{*}}=a_{i}^{*}.  \label{ca28}
\end{equation}

The next section shows the application of PCA method for knowledge discovery
in CAs. 
\Section{PCA and Cellular Automata \label{PCA-Cell}}

In this section we review the work presented in \cite{deniau94pca}. In this
reference, authors analyzed one-dimensional CAs using PCA. The key idea is
to consider binary patterns of a pre-defined size $l$ as inputs of the CAs.
It is considered the $256$ one-dimensional CA rules obtained for $r=1$ and $%
S=\left\{ 0,1\right\} $ in expression \ref{ca001}-\ref{ca01}. The output can
be collected in a Table, like Table \ref{pca00}, built for $l=5$.

\begin{table}[!htb]
\begin{center}
\begin{tabular}{llllll}
\hline
Patterns & $R_{0}$ & $R_{1}$ & ... & $R_{254}$ & $R_{255}$ \\ 
$00000$ & $000$ & $111$ & ... & $000$ & $111$ \\ 
$00001$ & $000$ & $110$ & ... & $001$ & $111$ \\ 
... & ... & ... & ... & ... & ... \\ 
$11110$ & $000$ & $000$ & ... & $111$ & $111$ \\ 
$11111$ & $000$ & $000$ & ... & $111$ & $111$ \\ \hline
\end{tabular}%
\end{center}
\caption{Table which rows are indexed by binary patterns and collumns by the
CA rules $R_{0}$, $R_{1}$, ..., $R_{255}$.}
\label{pca00}
\end{table}

Each row $j$ of Table \ref{pca00} is obtained through the application of the
rule $R_{j}$ (see expression \ref{regra-90-idex} for an example of rule indexation) Then, I/0
patterns are converted to cardinal numbers denoted by $f_{j}\left(
m_{i}\right) $, which means the cardinal number corresponding to the
application of the rule $j$ to the pattern $i$ ($i=0,1,...,31$ for Table 
\ref{pca00}). Thus, in general we get the matrix:

\begin{equation}
F=\left[ 
\begin{array}{ccc}
f_{11} & \ldots & f_{1p} \\ 
\vdots & \ddots & \vdots \\ 
f_{n1} & \ldots & f_{np}%
\end{array}
\right] ,  \label{pca01}
\end{equation}
where $f_{ij}=f_{j}\left( m_{i}\right) .$ The matrix $F$ is the data set to
be analyzed..

For mining knowledge in $F$ through PCA we should firstly to perform the
operation (translation) given by (\ref{ca18}). Thus, matrix $F$ is converted
to the following one:

\begin{equation}
X=\left[ 
\begin{array}{ccc}
x_{11} & \ldots & x_{1p} \\ 
\vdots & \ddots & \vdots \\ 
x_{n1} & \ldots & x_{np}%
\end{array}
\right] ,  \label{pca02}
\end{equation}
with: 
\begin{equation}
x_{ij}=f_{j}(m_{i})-E_{j},  \label{pca03}
\end{equation}
\begin{equation}
E_{j}=\frac{1}{n}\sum_{i=1}^{n}f_{j}(m_{i}).  \label{pca04}
\end{equation}
\newline
The matrix $X$ is of size $np$. In \cite{deniau94pca} columns $x_{1},\ldots
,x_{p}$ of $X$ are called variables while rows $e_{1},\ldots ,e_{n}$ are
called covariables. However, we must observe that space dimension is the
number of rules $(p)$ and the number of data vectors is the number of
patters $\left( n\right) .$ Thus, following section \ref{PCA}, we should
apply the PCA over the data set given by matrix $X^{T}$ in order to find out
the principal components of the covariables space. Besides, in \cite%
{deniau94pca} the norm in the covariables space is defined by:

\begin{equation}
M=diag\left( \frac{1}{S_{1}^{2}},\frac{1}{S_{2}^{2}},...,\frac{1}{S_{p}^{2}}%
\right) ,  \label{pca05}
\end{equation}
with:

\begin{equation}
S_{j}^{2}=\frac{1}{n^{2}}\left( n\sum_{i=1}^{n}x_{ij}^{2}-\left(
\sum_{i=1}^{n}x_{ij}^{{}}\right) ^{2}\right) =\frac{1}{n}\sum_{i=1}^{n}%
\left( x_{ij}-E_{j}\right) ^{2}.  \label{pca06}
\end{equation}

Following section \ref{PCA}, we must solve equation (\ref{ca27}) to find the
eigenvalues and then apply expression (\ref{ca28}) to get the eigenvectors
in the desired representation. The Table \ref{Table-Original} shows the
largest eigenvalues of this matrix for the listed pattern sizes.

\begin{table}[!htb]
\begin{center}
\begin{tabular}{|c|ccccccc|}
\hline
l & $\lambda _{1}$ & $\lambda _{2}$ & $\lambda _{3}$ & $\lambda _{4}$ & $%
\lambda _{5}$ & $\lambda _{6}$ & $\lambda _{7}$ \\ \hline
4 & 52.6802 & 48.2214 & 36.8869 & 36.8263 & 36.3134 & 24.4539 & 18.6179 \\ 
5 & 58.2575 & 50.9776 & 37.2301 & 37.0399 & 30.7382 & 21.7355 & 18.0214 \\ 
6 & 59.5952 & 51.6519 & 37.3406 & 37.1109 & 29.3769 & 21.0940 & 17.8305 \\ 
7 & 59.9260 & 51.8197 & 37.3696 & 37.1296 & 29.0383 & 20.9358 & 17.7811 \\ 
9 & 60.0290 & 51.8721 & 37.3788 & 37.1355 & 28.9325 & 20.8865 & 17.7656 \\ 
12 & 60.0358 & 51.8755 & 37.3794 & 37.1359 & 28.9256 & 20.8833 & 17.7645 \\ 
\hline
\end{tabular}%
\end{center}
\caption{Eigenvalues of the correlation matrix.}
\label{Table-Original}
\end{table}

The main result is that the eigenvalues from the seventh rank are
dramatically smaller in magnitude ($104$ times) than the first seven ones.
Such observation led authors of \cite{deniau94pca} towards the following
conjecture:

\textbf{Conjecture:}\textit{\ The rank of }$R$\textit{\ is }$7$\textit{\ and
does not depend on the size }$l$ \textit{of patterns being considered. When }%
$l$ is increased the \textit{eigenvalues tend to characteristic values
obtained for }$l=12.$

This is the main result presented in \cite{deniau94pca}. Next, we show our
results by applying the same analysis but introducing randomness in the CA
behavior.

\Section{PCA and Probabilistic Cellular Automata \label{PCA-ProbCell}}

In this section we report some experimental results obtained in the presence
of noise which randomly reverses the CA output values with probability $p$.
In the first experiment, reported on Table \ref{Table-size5}, we set $l=5$
and take some values for the probability $p$ and compute the PCA for the
generated matrix.

\begin{table}[!htb]
\begin{center}
\begin{tabular}{|l|l|l|l|l|}
\hline
& P=0.2 & P=0.4 & P=0.6 & P=0.8 \\ \hline
$\lambda _{1}$ & 30.993277 & 14.347278 & 27.368784 & 28.491445 \\ \hline
$\lambda _{2}$ & 17.127385 & 13.29037 & 16.960658 & 17.279648 \\ \hline
$\lambda _{3}$ & 13.541144 & 12.885112 & 14.959948 & 14.604408 \\ \hline
$\lambda _{4}$ & 13.180114 & 11.816202 & 13.281494 & 13.752977 \\ \hline
$\lambda _{5}$ & 12.388441 & 11.678309 & 11.357819 & 12.844465 \\ \hline
$\lambda _{6}$ & 10.902415 & 11.37782 & 10.857413 & 10.829726 \\ \hline
$\lambda _{7}$ & 10.293973 & 11.020907 & 9.959074 & 9.7779532 \\ \hline
$\lambda _{8}$ & 9.90735 & 10.419589 & 9.5077103 & 9.2701578 \\ \hline
$\lambda _{9}$ & 9.2822227 & 10.270046 & 9.3582306 & 8.8721786 \\ \hline
$\lambda _{10}$ & 8.7295329 & 9.8165154 & 8.8984414 & 8.8107682 \\ \hline
$\lambda _{11}$ & 8.529595 & 9.2160415 & 8.5837995 & 8.2016704 \\ \hline
$\lambda _{12}$ & 7.8843864 & 8.9368721 & 8.3232351 & 8.0601548 \\ \hline
$\lambda _{13}$ & 7.6179283 & 8.8871718 & 8.1230224 & 7.6299203 \\ \hline
$\lambda _{14}$ & 7.0733465 & 8.4223863 & 7.8908832 & 7.5448918 \\ \hline
$\lambda _{15}$ & 6.8634222 & 8.2353647 & 7.401953 & 7.3142038 \\ \hline
$\lambda _{16}$ & 6.7723605 & 7.9216589 & 7.2366057 & 7.066622 \\ \hline
$\lambda _{17}$ & 6.4657069 & 7.4769038 & 6.6688247 & 6.635226 \\ \hline
$\lambda _{18}$ & 6.3216585 & 7.0701685 & 6.190632 & 6.1886655 \\ \hline
$\lambda _{19}$ & 6.1603787 & 6.7891207 & 6.1204134 & 6.0738189 \\ \hline
$\lambda _{20}$ & 5.8044919 & 6.6787658 & 5.7734589 & 5.8293511 \\ \hline
$\lambda _{21}$ & 5.6959129 & 6.2790179 & 5.7684386 & 5.5555388 \\ \hline
$\lambda _{22}$ & 5.3941432 & 6.0714589 & 5.2569218 & 5.472738 \\ \hline
$\lambda _{23}$ & 4.9930157 & 5.8968197 & 4.9528161 & 5.2651065 \\ \hline
$\lambda _{24}$ & 4.799646 & 5.6406084 & 4.7506673 & 4.896581 \\ \hline
$\lambda _{25}$ & 4.5991479 & 5.5455414 & 4.5796108 & 4.7932072 \\ \hline
$\lambda _{26}$ & 4.3609133 & 5.2887732 & 4.5370786 & 4.7514249 \\ \hline
$\lambda _{27}$ & 4.1529641 & 5.1079093 & 4.3017415 & 4.2932052 \\ \hline
$\lambda _{28}$ & 3.9754383 & 4.9468391 & 4.0258767 & 3.7535244 \\ \hline
$\lambda _{29}$ & 3.7421531 & 4.826271 & 3.850906 & 3.6713829 \\ \hline
$\lambda _{30}$ & 3.3309643 & 4.362729 & 3.7734524 & 3.4443371 \\ \hline
$\lambda _{31}$ & 3.1165726 & 3.4774291 & 3.3800895 & 3.0247033 \\ \hline
$\lambda _{32}$ & 7.075E-15 & 5.587E-15 & 5.641E-15 & 7.235E-15 \\ 
\hline\hline
\end{tabular}%
\end{center}
\caption{Principal components for patterns with $l=5$ and noise with
probability distribution given by $p$.}
\label{Table-size5}
\end{table}

We observe that the number of principal components is $31$ for all
tests. When size patterns are increased to ($l=6,7,9,12$) we observe (see
Tables \ref{Table-size6-31} and \ref{Table-size6-64} for $l=6$ ) the same
behavior but the number of principal components increases to $%
63,127,254,254, $ respectively. In these tests, for a fixed pattern size,
the noise intensity ($p$ value) did not seem to play a considerable effect
in the number of principal components if $p>0.2$. However if $p=0.0$, we
know from the conjecture of section \ref{PCA-Cell} that this number is $7$
for all cases considered. What happens for $0.0<p<0.2$?

Such question must be considered in further works by the viewpoint of the KL
transform, following the procedure of section \ref{PCA} for a random field $%
\mathbf{u}\left( i,j\right) $ in order to have a complete answer.

\begin{table}[!htb]
\begin{center}
\begin{tabular}{|l|l|l|l|l|}
\hline
& P=0.2 & P=0.4 & P=0.6 & P=0.8 \\ \hline
$\lambda {1}$ & 25.490674 & 9.0159666 & 25.916423 & 26.284694 \\ \hline
$\lambda {2}$ & 14.10028 & 8.5855201 & 14.41566 & 14.102228 \\ \hline
$\lambda {3}$ & 11.104733 & 7.9801912 & 10.70382 & 10.688536 \\ \hline
$\lambda {4}$ & 10.790121 & 7.7267394 & 10.065372 & 9.7791048 \\ \hline
$\lambda {5}$ & 9.0958191 & 7.4831015 & 9.6810207 & 9.1538961 \\ \hline
$\lambda {6}$ & 7.2510569 & 7.3765707 & 7.638434 & 6.6054597 \\ \hline
$\lambda {7}$ & 6.8531309 & 7.0389042 & 6.8077442 & 6.4860439 \\ \hline
$\lambda {8}$ & 6.286914 & 6.9484678 & 6.2549327 & 6.1013626 \\ \hline
$\lambda {9}$ & 6.1443606 & 6.6158667 & 6.121154 & 5.9897572 \\ \hline
$\lambda {10}$ & 5.8704541 & 6.4728947 & 5.8079843 & 5.7110224 \\ \hline
$\lambda {11}$ & 5.6239779 & 6.3245388 & 5.6570288 & 5.3881715 \\ \hline
$\lambda {12}$ & 5.4920906 & 6.1992113 & 5.3781214 & 5.3309815 \\ \hline
$\lambda {13}$ & 5.2459503 & 5.9229922 & 5.2887514 & 5.1818583 \\ \hline
$\lambda {14}$ & 5.0507806 & 5.7271535 & 5.0157005 & 5.0733092 \\ \hline
$\lambda {15}$ & 4.8419224 & 5.5844553 & 4.9008587 & 4.9569268 \\ \hline
$\lambda {16}$ & 4.7432239 & 5.4868919 & 4.7443148 & 4.9045772 \\ \hline
$\lambda {17}$ & 4.5913664 & 5.4394466 & 4.646143 & 4.8375055 \\ \hline
$\lambda {18}$ & 4.5541085 & 5.2349427 & 4.519477 & 4.6544779 \\ \hline
$\lambda {19}$ & 4.3368357 & 5.1286 & 4.4405929 & 4.3767439 \\ \hline
$\lambda {20}$ & 4.289522 & 5.0877569 & 4.3612835 & 4.255288 \\ \hline
$\lambda {21}$ & 4.0221739 & 4.8696669 & 4.1463912 & 4.1147127 \\ \hline
$\lambda {22}$ & 4.0053207 & 4.7376638 & 4.0277528 & 3.9599816 \\ \hline
$\lambda {23}$ & 3.9195876 & 4.6631356 & 3.9294454 & 3.8759049 \\ \hline
$\lambda {24}$ & 3.8132963 & 4.5676061 & 3.8562459 & 3.7592389 \\ \hline
$\lambda {25}$ & 3.6929406 & 4.5192326 & 3.7291963 & 3.6175464 \\ \hline
$\lambda {26}$ & 3.6059328 & 4.3866647 & 3.6566283 & 3.5898727 \\ \hline
$\lambda {27}$ & 3.5245448 & 4.2816878 & 3.4551909 & 3.4851947 \\ \hline
$\lambda {28}$ & 3.4689985 & 4.1503676 & 3.4185097 & 3.4233569 \\ \hline
$\lambda {29}$ & 3.3229986 & 4.0343815 & 3.2894165 & 3.376657 \\ \hline
$\lambda {30}$ & 3.2214124 & 3.905456 & 3.2094177 & 3.247659 \\ \hline
$\lambda {31}$ & 3.1659706 & 3.7940953 & 3.1942609 & 3.2206595 \\ 
\hline\hline
\end{tabular}%
\end{center}
\caption{The first $31$ principal components for patterns with $l=6$ and
noise with probability distribution given by $p$.}
\label{Table-size6-31}
\end{table}

\begin{table}[tbh]
\begin{center}
\begin{tabular}{|l|l|l|l|l|}
\hline
& P=0.2 & P=0.4 & P=0.6 & P=0.8 \\ \hline
$\lambda {32}$ & 3.1278811 & 3.6850001 & 3.0605835 & 3.1566209 \\ \hline
$\lambda {33}$ & 2.9093472 & 3.5911018 & 2.9959418 & 2.9694106 \\ \hline
$\lambda {34}$ & 2.8851128 & 3.4984597 & 2.903988 & 2.9241726 \\ \hline
$\lambda {35}$ & 2.8361213 & 3.3609776 & 2.8339349 & 2.9066998 \\ \hline
$\lambda {36}$ & 2.7116889 & 3.2629616 & 2.7802321 & 2.8511868 \\ \hline
$\lambda {37}$ & 2.6410113 & 3.1449082 & 2.7448714 & 2.7909378 \\ \hline
$\lambda {38}$ & 2.62132 & 3.1060263 & 2.5446362 & 2.6124111 \\ \hline
$\lambda {39}$ & 2.5625247 & 3.0036165 & 2.4679447 & 2.591938 \\ \hline
$\lambda {40}$ & 2.4466011 & 2.9296411 & 2.4130114 & 2.5655322 \\ \hline
$\lambda {41}$ & 2.3957291 & 2.7868727 & 2.3229128 & 2.4194093 \\ \hline
$\lambda {42}$ & 2.3022031 & 2.729422 & 2.2821061 & 2.3069397 \\ \hline
$\lambda {43}$ & 2.1920726 & 2.717123 & 2.1595839 & 2.2493672 \\ \hline
$\lambda {44}$ & 2.1758946 & 2.603908 & 2.1373086 & 2.1920309 \\ \hline
$\lambda {45}$ & 2.0950959 & 2.4974377 & 2.0678771 & 2.1697002 \\ \hline
$\lambda {46}$ & 2.0489445 & 2.3755221 & 1.9724868 & 2.0876546 \\ \hline
$\lambda {47}$ & 2.020732 & 2.3495888 & 1.9516421 & 2.0184278 \\ \hline
$\lambda {48}$ & 1.9777077 & 2.2768661 & 1.8642426 & 1.9885042 \\ \hline
$\lambda {49}$ & 1.8766222 & 2.1822128 & 1.8066548 & 1.8790247 \\ \hline
$\lambda {50}$ & 1.7943956 & 2.1355595 & 1.758534 & 1.8591266 \\ \hline
$\lambda {51}$ & 1.7171865 & 2.0430028 & 1.6877656 & 1.7836542 \\ \hline
$\lambda {52}$ & 1.6253386 & 2.022733 & 1.6355071 & 1.6965591 \\ \hline
$\lambda {53}$ & 1.5259488 & 1.9193735 & 1.6129515 & 1.6380111 \\ \hline
$\lambda {54}$ & 1.4965356 & 1.807634 & 1.4445091 & 1.5871297 \\ \hline
$\lambda {55}$ & 1.4114911 & 1.7544899 & 1.3778514 & 1.5641227 \\ \hline
$\lambda {56}$ & 1.381066 & 1.7005373 & 1.3387275 & 1.5194184 \\ \hline
$\lambda {57}$ & 1.3691854 & 1.6217295 & 1.2709127 & 1.3652494 \\ \hline
$\lambda {58}$ & 1.2926625 & 1.4695063 & 1.2284464 & 1.2780083 \\ \hline
$\lambda {59}$ & 1.1645091 & 1.4539369 & 1.1474655 & 1.2135036 \\ \hline
$\lambda {60}$ & 1.1101623 & 1.3227765 & 1.1182555 & 1.1747427 \\ \hline
$\lambda {61}$ & 0.9979225 & 1.1870755 & 1.0455127 & 1.120168 \\ \hline
$\lambda {62}$ & 0.9866537 & 1.1730686 & 0.9469073 & 1.0335351 \\ \hline
$\lambda {63}$ & 0.7798336 & 0.9967607 & 0.7994225 & 0.9540748 \\ \hline
$\lambda {64}$ & 4.890E-15 & 3.123E-15 & 8.151E-15 & 6.998E-15 \\ 
\hline\hline
\end{tabular}%
\end{center}
\caption{The last $33$ principal components for patterns with $l=6$ and
noise with probability distribution given by $p$.}
\label{Table-size6-64}
\end{table}

However, an interesting points is that these question resemble the problem
of studying the influence of control parameters in continuous dynamical
systems \cite{Haken1987}. With such parameters we can control the influence
of factors like temperature, viscosity, irradiation, etc. Those systems 
can
be analyzed through stability analysis \cite{Iooss1997,Haken1987},
bifurcation and catastrophe theory \cite{Iooss1997,Postol1978} and
perturbation \cite{Haken1987}. In that context, there may be critical 
values
for the parameters, in the sense that sudden changes happen near them. As an
example, let us consider a simple dynamical system:

\begin{eqnarray}
\frac{dx}{dt} &=&\lambda x+y  \label{dyn00} \\
\frac{dy}{dt} &=&x+\lambda y  \notag
\end{eqnarray}%
where $\lambda $ is a real parameter. According to the theory of ordinary
differential equations \cite{Haken1987}, the qualitative analysis of this
system may be done through the analysis of the eigenvalues/eigenvectors of
the matrix of the above system:

\begin{equation}
A=\left[ 
\begin{array}{ll}
\lambda & 1 \\ 
1 & \lambda%
\end{array}
\right] .  \label{dyn01}
\end{equation}

The eigenvalues are given by:

\begin{equation}
\alpha _{1}=\lambda +1,\quad \alpha _{2}=\lambda -1.  \label{dyn02}
\end{equation}

We observe that the value $\lambda =1$ is a critical one because, for $%
\lambda >1$ the origin $\left( 0,0\right) $ is an attractor but for $\lambda
<1$ we observe a saddle point. Thus, we have a \textit{jump}, that is, a
sudden change in the system behavior, for $\lambda =1.$

Cellular automata are discrete dynamical systems for which the probability $%
p $ could be seem as a parameter that ranges in $\left[ 0,1\right] $. Would
there be critical values in this case? If the answer is ''yes'' which
property suddenly changes? Could it be the number of principal components?

These questions and the mathematical theory necessary to perform
such analysis are points that we shall consider in further works.

\Section{Final Comments \label{Final}}

In this paper we review the application of PCA for cellular automata
analysis. We follow the work presented in \cite{deniau94pca} but in the
presence of noise which randomly reverses the CA output values with
probability $p$. We observe that, for a fixed pattern size, the noise
intensity ($p$ value) did not seem to play a considerable effect in 
the
number of principal components if $p>0.2$. The observed (and expected)
influence is that the number of principal components increases. For example,
if $l=5$ and $p=0.0$, the main result of \cite{deniau94pca} says that this
number is $7$ while for $p\in \left\{ 0.2,0.6,0.8\right\} $ this number
increases to $31$. The obvious question is that what happens for $0.0<p<0.2$?

This and others questions about parameter sensitivity analysis for cellular
automata must be answered in further works.

\Section{ Acknowledgments}

We would like to acknowledge CNPq for the financial support for this work.

\bibliographystyle{latex8}
\bibliography{ca-ref-gilson}

\end{document}